# Optoelectronic Sensitization of Carbon Nanotubes by CdTe Nanocrystals

B. Zebli,[1] H.A. Vieyra,[1] I. Carmeli,[2] A. Hartschuh,[3] J.P. Kotthaus,[1] A.W. Holleitner[1,4,*]

(1) Department für Physik and Center for NanoScience, Ludwig-Maximilians-Universität, Geschwister-Scholl-Platz 1, 80539 München, Germany.

(2) Department of Chemistry and Biochemistry, Tel-Aviv University, Tel-Aviv 69978, Israel.

(3) Department für Chemie, Physikalische Chemie, Butenandtstr. 5-13 E 81377 Munich, Germany.

(4) Walter-Schottky Institut, Technische Universität München, Am Coulombwall 3, 85748 Garching, Germany.

We investigate the photoconductance of single-walled carbon nanotube-nanocrystal-hybrids. The nanocrystals are bound to the nanotubes via molecular recognition. We find that the photoconductance of the hybrids can be adjusted by the absorption characteristics of the nanocrystals. In addition, the photoconductance of the hybrids surprisingly exhibits a slow time constant of about 1 ms after excitation of the nanocrystals. The data are consistent with a bolometrically induced current increase in the nanotubes caused by photon absorption in the nanocrystals.

*e-mail: holleitner@wsi.tum.de

PACS number(s): 73.22.-f, 73.50.Pz, 65.80.+n,

## I. Introduction:

Carbon nanotubes (CNTs) have emerged as promising building blocks of nanoscale optoelectronic devices [1]-[6]. The functionalization of CNTs by chemical modifications holds interesting prospects in various fields, such as the fabrication of hybrid bioorganic nanosystems [7],[8]. In particular, it has been demonstrated how to bind CNTs to single molecules [9], photosynthetic proteins [10], graphite beads [11], and nanocrystals [12]-[17]. The adjustable optical properties of colloidal semiconductor nanocrystals make them very suitable for optoelectronic devices such as solar cells [18]. Here, we demonstrate that the photoconductance of CNTs can be photo-sensitized by the absorption characteristics of chemically attached CdTe nanocrystals. In order to explore this optoelectronic sensitization effect, we measure the photoconductance of nanotube-nanocrystal-hybrids as a function of the photon energy as well as the polarization and modulation frequency of the incident laser light. We describe an optoelectronic sensitization effect with a slow time constant of about 1 ms, when the nanocrystals of the hybrids are optically excited. We interpret the slow time constant such that if the nanocrystals are optically excited, the lattice temperature of the CNTs is raised. Hereby, the current across the CNTs is increased [6]. This indirect bolometric effect allows photo-sensitizing CNTs by semiconductor nanocrystals. We find good agreement between the experimental data and a model which includes both heat and Förster resonance energy transfer between the nanocrystals and the nanotubes. If only the CNTs are excited, we find a fast photoconductance time constant, which is consistent with electron-hole effects, as recently reported for pure CNTs [4],[5],[22].

## II. Material Synthesis and Characterization:

Starting point of the synthesis of nanocrystal-nanotube-hybrids are purified single-walled CNTs with carboxyl groups at the side-walls and the tips [10],[23]. The hybrids are then synthesized by applying the carboxylated CNTs successively to 1-ethyl-3-(3-dimethylaminopropyl)-carbodiimide-hydrochloride (EDC) and N-hydroxysulfo-succinimide (Sulfo-NHS) to form a semi-stable NHS-ester bound to the CNT, to biotin-PEO-amine, and to streptavidin-coated CdTe nanocrystals [see steps I, II, III, and IV in Fig. 1]; following ref. [10], [16], and [24]. In the transmission electron microscope (TEM) image of Fig. 2(a), the purified CNTs appear as thin lines, wrapped by a tissue consisting of the utilized dried solvent TritonX - a non-ionic surfactant. Fig. 2(b) shows a TEM image of a hybrid with CdTe nanocrystals. The nanocrystals appear as dark dots with a size of about 10 nm. Importantly, the nanocrystals are always located on the CNTs within the solvent coating. Because the samples are filtered several times before they are assembled onto the TEM carriers, we interpret the observation such that the nanocrystals are bound to the CNTs.

In order to determine the diameter of the CNTs, we measure the fluorescence and Raman signal of the hybrid material deposited on a glass coverslide excited at $E_{PHOTON}$ = 1.96 eV (632.8 nm) at room temperature [Fig. 3(a)]. The sharp peaks refer to the Raman signal of the CNTs (black triangles). Raman spectra of the ensemble show a single radial-breathing mode contribution at 200 cm$^{-1}$ [Fig. 3(b)]. Hereby, we can determine the radius of metallic CNTs electronically resonant at $M_{11}$ ~ 1.96 eV to be $R_{CNTS}$ = (0.63 ± 0.04) nm [25]. Assuming the same radius for semiconducting CNTs of the ensemble, it translates to an energy of $E_{11}$ = (0.66 ± 0.05) eV and $E_{22}$ = (1.325 ± 0.125) eV for the first and second semiconducting interband transition [3],[25]. Fig. 3(a) also demonstrates that the nanocrystals (NC) within the hybrid material show a fluorescence emission energy $E_{NC}$ centered at 1.59 eV. The fluorescence signal of the bound nanocrystals in the hybrid material is consistent with the emission curve of unbound streptavidin-coated CdTe nanocrystals [Fig. 3(c)]. Fig. 3(c) depicts the extinction curve of the unbound streptavidin-coated CdTe nanocrystals as a function of $E_{PHOTON}$, which increases for $E_{PHOTON} \geq E_{NC}$. Fig. 3(d) shows the extinction curves for the linker molecules biotin and streptavidin solved in water. As a result, we can assume that both linker molecules do not absorb photons in the range of 1.31 eV $\leq E_{PHOTON} \leq$ 1.77 eV, in which the photoconductance measurements are performed.

**III. Experiment:**

In the following, we present photoconductance and conductance measurements on ensemble samples of each of the four steps I, II, III, and IV (as defined in Fig. 1) and on four single nanotube-nanocrystal-hybrid samples called A, B, C, and D . The ensemble samples are electronically contacted by depositing a drop of an aqueous solution containing the nanotube-nanocrystal-hybrids onto an insulating $SiO_2$ substrate with lithographically predefined gold contacts on top. The electronic contacts for the ensembles have a distance of 3-5 µm and a height of ~100 nm, and they are electronically bridged by the ensembles after the water has been dried out; forming a two-terminal circuit. The single hybrids are contacted by source and drain contacts made of Pd utilizing e-beam lithography [16] [e.g. see sample A in Fig. 4(a)]. All photoconductance measurements are carried out in a helium continuous-flow cryostat at a vacuum of <$10^{-3}$ mbar in combination with a titanium:sapphire laser. The laser is continuously tunable between $E_{PHOTON}$ = 1.31 eV (950 nm) and 1.77 eV (700 nm). The light is linearly polarized, and the polarization angle $\varphi$ can be controlled with respect to the orientation $\hat{w}$ of the single hybrid [see tripod in Fig. 4(a)]. We measure the photoconductance $G(E_{PHOTON}, \hat{E}, f_{CHOP})$ of the hybrid nanostructures as a function of the photon energy $E_{PHOTON}$, the polarization $\hat{E}$ of the photon, and the chopper frequency $f_{CHOP}$. To this end, a bias voltage $V_{SD}$ is applied across the source-drain electrodes in Fig. 4(a), while the laser is focused onto the hybrid nanostructure. The photoconductance $G_{PHOTON}(E_{PHOTON}, \hat{E}, f_{CHOP}) = [I_{ON}(E_{PHOTON}, \hat{E}, f_{CHOP}) - I_{OFF}]/V_{SD}$ across the sample with the laser being in the "on" or "off" state, respectively, is amplified by a current-voltage converter and detected with a lock-in amplifier using the chopper signal as a reference. Fig. 4(b) depicts a typical photoconductance signal $G_{PHOTON}$ for an ensemble sample of step IV as a function of the laboratory time for $E_{PHOTON}$ = 1.61 eV (open circles) and $E_{PHOTON}$ = 1.41 eV (black circles). The data will be discussed in section IV and V. We would like to note that the signal as in Fig. 4(b) is symmetric with respect to $V_{SD}$ [data not shown], and that it does not exhibit a finite value at zero bias. Therefore, we report on photoconductance phenomena in the nanocrystal-nanotube hybrids, but not on photocurrent or -voltage processes.

The data Fig. 4(c) represent a typical current-voltage characteristic of the single nanotube-nanocrystal-hybrid sample B without laser excitation for a bath temperature in the range of 5 K ≤ $T_{BATH}$ ≤ 100 K. We typically observe a temperature-dependent non-linear current-voltage characteristic, which can be translated into a temperature gradient of $\Delta(I_{SD}/V_{SD}) / \Delta T_{BATH}$ ≈ +3 nS /K at $V_{SD}$ = 20 mV. We interpret the temperature dependence of the current-voltage characteristic to be caused either by numerous defects of the functionalized CNTs, by slightly non-ohmic and non-symmetric source/drain contacts, or by Coulomb blockade effects [16]. As will be shown in section V, the temperature gradient is important for the final interpretation of the photoconductance properties of the nanotube-nanocrystal-hybrids. However, the interpretation is independent of the gradient's cause.

**IV. Results:**

Generally, the photoconductance across CNTs shows resonances whose energies coincide with the intersubband transitions $E_{11}$ and $E_{22}$ in single-walled CNTs [4],[5]. We verify this prediction by photoconductance measurements on ensemble samples, containing CNTs which are functionalized according to steps I, II, and III in Fig. 5(a)-(c). For such samples, the photoconductance decreases by about 40-50 % for $E_{PHOTON}$ = 1.66 eV compared to 1.38 eV [Fig. 5(a)-(c)] [26]. This trend reproduces the absorption spectrum of semiconducting CNTs [1] and, the data are consistent with an $E_{22}$ = (1.325 ± 0.125) eV found from Raman spectroscopy. We would like to note that the appearance of a maximum in Fig. 5(c) is likely to be an artefact; reflecting the variation of the carbon nanotubes' composition from ensemble sample to sample. As can be seen in Fig. 3(d), biotin only absorbs light above $E_{PHOTON} \approx$ 5 eV. Hereby, the functionalization of the CNTs does not explain the maximum in Fig. 5(c). Most importantly, Fig. 5(d) demonstrates that the nanotube-nanocrystal-hybrids [step IV in Fig. 1] opposes the described trend, such that the photoconductance increases by about 10-20% for $E_{PHOTON}$ = 1.66 eV compared to 1.38 eV. Hereby, the spectral dependence of the photoconductance in Fig. 5(d) strongly mimics the absorption characteristics of the CdTe nanocrystals in Fig. 3(c).

Fig. 6 shows the photoconductance of further ensemble samples - which are functionalized according to steps I, II, III, and IV - as a function of $f_{CHOP}$. Up to a chopper frequency of ~3.5 kHz and within the error bars, the photoconductance of samples functionalized according to steps I, II, and III is constant for all photon energies. For the nanotube-nanocrystal-hybrids [step IV in Fig. 1] we find the photoconductance to be independent of the chopper frequency for $f_{CHOP} \leq 3.5$ kHz and $E_{PHOTON} \leq E_{NC}$ [Fig. 6(d)]. For $E_{PHOTON} \geq E_{NC}$, however, we observe a decrease of the photoconductance for an increasing $f_{CHOP}$ and $f_{CHOP} \leq 1$ kHz [dotted oval in Fig. 6(d)]. For a larger chopper frequency, we again detect a rather constant photoconductance value. The line in Fig. 6(d) is a fit to following equation, describing a frequency domain response measurement [27]:

$$G(f_{CHOP}) = (G_0 - G_{OFF-SET})/\sqrt{1 + (2\pi f_{CHOP} \cdot \tau_0)^2} + G_{OFF-SET}, \qquad (1)$$

with $G_0$ the photoconductance at $f_{CHOP} = 0$ Hz, $\tau_0 \cong (1.2 \pm 0.7)$ ms the response time of the photoconductance, and $G_{OFF-SET}$ a photoconductance off-set for large $f_{CHOP}$. In section V, we will interpret the response time to reflect a bolometric increase of the lattice temperature of the CNTs, which is induced by photon absorption in the CdTe nanocrystals.

In Fig. 7, we show photoconductance data of single nanotube-nanocrystal-hybrid samples A, B, C, and D. Fig. 7(a) depicts the photoconductance of sample A as a function of the polarization angle $\varphi$ at $E_{PHOTON} = 1.38$ eV at $T_{BATH} = 4.5$ K. For all photon energies, we observe that the photoconductance exhibits a maximum when the light is polarized along the direction of the hybrid axis $\hat{w}$. The observation is consistent with recent measurements on single semiconducting and metallic CNTs, which act as submicron "antennas" [5],[29]. In the present case of nanotube-nanocrystal-hybrids, we observe that the photoconductance at a perpendicular orientation of the polarization increases for $E_{PHOTON} \geq E_{NC}$ [demonstrated for sample C in Fig. 7(b)]. We interpret the second observation such that the nanocrystals contribute to the hybrid's photoconductance independently of the orientation of the photon polarization. Hereby, the photoconductance of the single hybrid reveals the absorption characteristic of the nanocrystals. This fingerprint of the nanocrystals is corroborated by chopper frequency dependent measurements. At $E_{PHOTON} \leq E_{NC}$ [in Fig. 7(c)], we find the photoconductance of sample C to be independent of the chopper frequency up to $f_{CHOP} \approx 3.5$ kHz. For $E_{PHOTON} \geq E_{NC}$ [Fig. 7(d)], the photoconductance of sample C decreases for increasing chopper frequency for perpendicular and parallel polarization of the photon. Fitting the curves to Eq. (1) gives a response time of about $\tau_0 \cong (0.3 \pm 0.1)$ ms and $G_0 \approx 12.5$ nS. We find that $\tau_0$ varies from sample to sample. For instance, fitting the data of sample B in Fig. 7(e) to Eq. (1) gives $\tau_0 \cong (2.0 \pm 0.2)$ ms and $G_0 \approx 6.5$ nS. We can estimate the average response time of samples A, B, C, and D to be $\tilde{\tau}_0 \cong 1.1$ ms in the range between 0.3 ms and 2 ms. The data in Fig. 7(e) also corroborate the measurements on the ensemble sample in Fig. 6(d), such that there is a finite photoconductance signal for a large frequency. Finally, we demonstrate that the observation of an enhanced photoconductance at large photon energies is robust even at room temperature. To this end, Fig. 7(f) displays the photoconductance of the single hybrid sample D as a function of $E_{PHOTON}$ at an arbitrary polarization of the exciting photons at room temperature. Again, the photoconductance of sample D increases for $E_{PHOTON} \geq E_{NC}$.

Generally, we observe that the described effects depend on the number of CdTe nanocrystals attached to the CNTs. Devices with a single hybrid such in Fig. 4(a) enable us to estimate a lower threshold to be approximately $N_{NC} \sim 8$ µm$^{-1}$ for the number of nanocrystals per length of a CNT necessary to detect the optoelectronic sensitization effect. This finding is consistent with recent reports on the absorption cross section of single CNTs with about 1 µm length under non-resonant excitation ($\sigma_{CNT} \sim 10^{-15}$ cm$^2$) and the integrated cross section of all attached CdTe nanocrystals per 1 µm length ($\sigma \sim N_{NC} \cdot \sigma_{NC} \sim 8 \cdot 10^{-15}$ cm$^2 > \sigma_{CNT}$) [29].

### V. Discussion:

There are several processes which can alter the photoconductance of the nanotube-nanocrystal-hybrids, such as photodesorption-effects on the surface of the CNTs [21], the effect of Schottky contacts between the CNTs and the metal electrodes [28], electron-hole effects within the CNTs [4][5], bolometric effects [6], and charge as well as energy transfer processes within the hybrids [32]. All of our samples are measured in vacuum ($p < 1 \times 10^{-3}$ mbar). Hereby, we can rule out photodesorption effects, where the laser excitation induces an oxygen-desorption of the dopant oxygen from the side-walls of the CNTs [21]. We can also rule out the effect of a Schottky barrier between the CNTs and the metal contacts as the dominating optoelectronic effect, because we do not detect any off-set in $V_{SD}$ or $I_{SD}$ of the photoconductance (data not shown) [28]. However, we do observe the electron-hole effects in the CNTs as discussed in conjunction with Fig. 5(a)-(c).

In Fig. 8, we sketch different heat transfer processes and a Förster resonance energy transfer within the nanotube-nanocrystal-hybrids when the CdTe nanocrystals are optically excited. It is well reported in literature that electron relaxation processes in the CNTs occur on a fast time-scale of picoseconds [22], while processes related to the phonon bath can sustain on a longer time-scale of milliseconds [6]. The measurements in Fig. 6(d), 7(d), and 7(e) demonstrate that we observe such a long time-scale with an average response time of $\tilde{\tau}_0 \cong 1.1$ ms for $E_{\text{PHOTON}} \geq E_{\text{NC}}$. Generally, we can estimate the maximum temperature increase $\Delta T_{\text{NC}}^{\text{MAX}}$ of a CdTe nanocrystal after laser excitation by comparing the absorbed laser power to the specific heat of the CdTe nanocrystals with a heat coupling to their surroundings, which is characterized by the typical optoelectronic response time of the hybrids $\tilde{\tau}_0$ as [30]

$$\Delta T_{\text{NC}}^{\text{MAX}} = \frac{P_{\text{abs}} \tilde{\tau}_0}{C_{\text{NC}}} \approx \frac{0.6 \cdot \sigma_{\text{NC}} I_{\text{in}} \tilde{\tau}_0}{C_{\text{NC}}}, \qquad (2)$$

with $P_{\text{abs}} = \sigma_{\text{NC}} I_{\text{in}}$ the absorbed laser power being transformed into heat in the nanocrystals with the absorption cross section of the nanocrystals $\sigma_{\text{NC}} = 10^{-15} cm^2$ [29], $I_{\text{in}} = 1.9 \, W/cm^2$ the incident light intensity, and $C_{\text{NC}}$ the corresponding specific heat. The factor of 0.6 accounts for the fluorescence quantum yield in CdTe nanocrystals of up to 40 % [35]. Following ref.[36]-[38],[40], we estimate the specific heat of the CdTe nanocrystals to be $C_{\text{NC}} \approx 1.6 \, J \cdot mol^{-1} K^{-1}$ at a temperature of 7 K. Inserting the average of the response time $\tilde{\tau}_0 \cong 1.1$ ms into Eq. (1), the temperature increase for a nanocrystal can be evaluated to be $\Delta T_{\text{NC}}^{\text{MAX}} \approx 16 \, K$ for experimental parameters such as in Fig. 7(d) [39]. We would like to note, that the optoelectronic response time of the hybrids $\tilde{\tau}_0$ is an upper limit of the phonon relaxation time of the phonon bath in the nanocrystals. Hereby,

Eq. (2) gives an upper limit of $\Delta T_{NC}^{MAX}$. The minimum temperature increase $\Delta T_{NC}^{MIN}$ can be estimated as [33]

$$\Delta T_{NC}^{MIN} = \frac{0.6 \cdot \sigma_{NC} I_{in}}{4\pi k_{SiO2} R_{NC}}, \qquad (3)$$

with $R_{NC}$ the nanocrystal radius and $k_{SiO2}$ the thermal conductivity between the nanocrystal and the substrate made of SiO$_2$ ($k_{SiO2} \approx 1$ mW/Kcm at 4 K [34]). We evaluate $\Delta T_{NC}^{MIN}$ to be about 0.2 µK for the experimental parameters as for Fig. 7(d). In Eq. (3), the temperature increase in the CdTe nanocrystals is underestimated for three reasons. First, in Eq. (3) only the substrate's thermal conductivity is considered, whereas the porous linker polymer can be assumed to have a smaller thermal conductivity than $k_{SiO2}$. Second, Eq. (3) neglects the acoustic impedance mismatch between the nanocrystals and their environment, i.e. it assumes a perfect thermal contact between nanocrystal and substrate. Third, only a section of the nanocrystals touches the substrate or the CNT. All three arguments make a further heat accumulation in the CdTe nanocrystal plausible.

Based upon the temperature coefficient $\Delta(I_{SD}/V_{SD}) / \Delta T_{BATH} \approx +3$ nS /K and a typical photoconductance increase of $2(G_0 - G_{OFF-SET}) \approx 11$ nS of sample B [Fig. 4(c) and Fig. 7], we can estimate an average increase of the lattice temperature in the CNT of sample B to be $\Delta T_{LATTICE}^{CNTs} \sim (G_0 / [\Delta(I_{SD}/V_{SD}) / \Delta T_{BATH}]) \approx 4$ K. This value is consistent with the lower and upper boundary of the estimated temperature in the nanocrystals. In turn, the decrease of the photoconductance for increasing $f_{CHOP}$ and $f_{CHOP} < 1$ kHz in Fig. 6(d), 7(d), and 7(e) reveals a relaxation time of the photoconductance signal in the order of a millisecond, which we interpret to be due to an indirect bolometric effect in the hybrids; i.e. photons are absorbed in the nanocrystals, and the conductance in the CNT is bolometrically increased. In this interpretation, the variation of the relaxation time from sample to sample reflects the different heat coupling of the specific hybrid to the surrounding. As already mentioned in section III, the temperature dependence of the current-voltage characteristic of the nanotube-nanocrystals-hybrids is caused either by the numerous defects of the functionalized CNTs, by slightly non-ohmic and non-symmetric source/drain contacts, or by Coulomb blockade effects [16]. The gradient's cause, however, does not influence the interpretation of a bolometrically increased conductance of the CNTs.

Fig. 8 further highlights a possible Förster resonance energy transfer (FRET) process between the optically excited nanocrystals and the CNTs. The transfer rate of excitons $\gamma_{FRET}$ from the nanocrystals to the CNTs can be modelled as [32]:

$$\gamma_{FRET} = \frac{9\pi}{32\hbar} \frac{R_{CNT}^2}{d^5} \left(\frac{ed_{exc}}{\varepsilon_{eff}}\right)^2 \cdot \mathrm{Im}\,\varepsilon_{CNT}, \qquad (4)$$

with $R_{CNT}$ the radius of the CNTs, $\varepsilon_{eff}$ the effective dielectric constant for the nanocrystal, $\mathrm{Im}\,\varepsilon_{CNT}$ the imaginary part of the dielectric constant of the CNTs, and $d_{exc}$ the dipole moment of the exciton. The center-to-center distance $d$ between the nanocrystals and the CNTs can be estimated to be in the range of $d_{MIN} \leq d \leq d_{MAX}$, with $d_{MIN} = R_{NC} + R_{CNT} + d_{STREPTAVIDIN} \sim (5 + 0.6 + 5)$ nm $= 10.6$ nm and $d_{MAX} = R_{NC} + R_{CNT} + d_{STREPTAVIDIN} + R_{LINKER} \sim (5 + 0.6 + 5 + 2)$ nm $= 12.6$ nm, $d_{STREPTAVIDIN}$ the thickness of the streptaviding coating of the nanocrystals, and $R_{LINKER}$ the maximum length of the linker molecules [as in Fig. 1(d)]. Following ref. [32] $d_{exc}$ and $\varepsilon_{eff}$ can be assumed to be $d_{exc} = 0.8$ Angstrom and $\varepsilon_{eff} = (2\varepsilon_0 + \varepsilon_{NC})/3$, with $\varepsilon_{NC} = 7.2$ the dielectric constant of the CdTe nanocrystals, and $\varepsilon_0 = 8.85 \cdot 10^{-12}$ C$^2$/Jm. In turn, we find $10^6$ s$^{-1} \leq \gamma_{FRET} \leq 10^7$ s$^{-1}$. For a typical exciton lifetime in CdTe nanoparticles of $\tau_{exc}$ = (10 - 30) ns [20], we can evaluate the Förster radius $R_0 = d \cdot \sqrt[5]{\gamma \tau_{exc}}$ to be in the range of 6.8 nm ≤ $R_0$ ≤ 8.5 nm. As a result, we can estimate the FRET efficiency $E_{FRET} = [1+(d/R_0)^5]^{-1}$ to be in the range of 0.04 ≤ $E_{FRET}$ ≤ 0.25. The FRET efficiency describes the fraction of energy transfer events, which occur between the CdTe nanocrystals and the CNTs, per excitation event in the CdTe nanocrystals. Such transferred excitations can give rise to a heated electron bath in the CNTs (Fig. 8). Because the electron temperature in the CNTs reaches equilibrium within picoseconds [22], an electronic bolometric effect should be limited by $\gamma_{FRET}$. As a result, we expect an off-set value of the photoconductance of the nanotube-nanocrystal-hybrids for a higher chopper frequency and $E_{PHOTON} > E_{NC}$. Indeed, the data in Fig. 6(d) and 7(e) exhibit such a finite photoconductance for $f_{CHOP}$ > 1 kHz. In addition, we would like to note that an elevated temperature of the electron bath in the CNTs can increase the temperature of the phonon bath in the CNTs (Fig. 8). Such a process again gives rise to the already discussed phonon induced bolometric photoconductance with a slow time constant.

Finally, we would like to note that our work describes processes which are different with respect to recent work by Juarez et al [17]. Juarez et al. reported on a decrease of the conductance of nanotube-nanocrystal-hybrids, when the nanocrystals are optically excited. This decrease occurs on a time-scale of several seconds. Juarez et al. suggested that electrons could tunnel from the nanocrystals to the CNTs such that positively charged nanocrystals remain. The charge compensation of the remaining holes via reactions with the environment would happen on a slow time-scale of several seconds [17]. As can be seen in Fig. 4(b), and according to our estimates of charge carrier tunnel probabilities in the order of $10^{-48}$, such processes do not dominate the photoconductance dynamics which we describe in this manuscript .

In summary, we present fluorescence, transport, and photoconductance measurements on nanotube-nanocrystal-hybrids. The nanocrystals are bound to the CNTs via molecular recognition. We find that the photoconductance of the hybrids can be adjusted by the absorption characteristics of the nanocrystals. We interpret the data such that photons are absorbed in the nanocrystals, and in turn, both the lattice and the electron temperature in the CNTs are locally increased. Such an increase enhances the conductance of the CNTs. We discuss a possible Förster resonance energy transfer process as well as heat transfer processes within the nanotube-nanocrystal-hybrids and find reasonable agreement between our model calculation and the experimental data.

We deeply thank A.O. Govorov for extensive discussions as well as M. Döblinger and S. Manus for technical support. In addition, we gratefully acknowledge financial support by the DFG-SFB 486 and the DFG-project HO 3324/2, the Center for NanoScience (CeNS) in Munich, and the German excellence initiative via the cluster "Nanosystems Initiative Munich (NIM)", and the program LMUexcellent, and the program LMUInnovativ FuNS.

FIG 1: Four-step chemical scheme to build nanotube-nanocrystal-hybrid. (a) Step I: after an acid treatment, carbon nanotubes (CNTs) exhibit carboxyl groups at the side-walls and the tips. (b) Step II: a semi-stable NHS-ester is bound to the CNTs. (c) Step III: biotinylized CNTs are obtained by binding biotin-PEO-amine to EDC. (d) Step IV: streptavidin-functionalized CdTe nanocrystals (NCs) bind to the biotin-functionalized CNTs. Graphs are not to scale (diameter of nanocrystals ~10 nm, diameter of CNTs ~1.2 nm, thickness of the streptavidin layer ~5 nm).

FIG. 2: (a) Transmission electron microscope (TEM) image of purified CNTs. (c) TEM image of nanotube-nanocrystal-hybrids.

FIG. 3. (a) Emission spectrum of dry nanotube-nanocrystal-hybrids excited at $E_{PHOTON}$ = 1.96 eV at room temperature. Center maximum refers to fluorescence of the nanocrystals. CNTs produce a Raman signal as sharp Raman spikes (black triangles). (b) Corresponding Raman signal at room temperature. (c) Fluorescence and exctinction spectrum of streptavidin-coated, unbound colloidal nanocrystals with a fluorescence maximum at $E_{NC}$ = 1.59 eV (white triangle) at room temperature. (d) Extinction spectrum of biotin and streptavidin solved in water at room temperature.

FIG. 4. (a) Atomic force microscope (AFM) image of an individual nanotube-nanocrystal-hybrid contacted by source and drain electrodes (sample A). Inset: incident light is linearly polarized along direction $\hat{E}$. Polarization angle $\varphi$ can be controlled with respect to the orientation $\hat{w}$ of the hybrid. (b) Typical photoconductance measurement of an ensemble sample made of nanotube-nanocrystal-hybrids as a function of the laboratory time at $E_{PHOTON}$ = 1.61 eV (open circles) and $E_{PHOTON}$ = 1.41 eV (black circles). (c) Typical current-voltage characteristic of a single nanotube-nanocrystal-hybrid without any laser excitation at a bath temperature $T_{BATH}$ of 5, 10, 15, 20, 30, 40, 50, 75, and 100 K.

FIG. 5: (a), (b), (c), and (d): Normalized photoconductance of ensemble samples of steps I, II, III, and IV (as in Fig. 2) as a function of $E_{PHOTON}$ (at room temperature, $P$ = 300 µW, $V_{SD}$ = 300 mV, $f_{CHOP}$ = 400 Hz, pulsed laser excitation).

FIG. 6: (a), (b), (c), and (d): Normalized photoconductance of different ensemble samples as compared to Fig. 5 from the preparation steps I, II, III, and IV as a function of $f_{CHOP}$ at $E_{PHOTON}$ = 1.38 eV, and 1.72 eV (at room temperature, $P$ = 1.3 mW, $V_{SD}$ = 70 mV, cw laser excitation).

FIG. 7: Photoconductance data of individual nanotube-nanocrystal-hybrids. (a) Normalized photoconductance of sample A as a function of $\varphi$ ($f_{CHOP}$ = 315 Hz, $E_{PHOTON}$ = 1.38 eV, $P$ = 300 nW, cw-laser excitation). (b) Normalized photoconductance of sample C as function of $E_{PHOTON}$ at perpendicular (circles) and parallel (open circles) polarization (at $T_{BATH}$ = 4.5 K, $f_{CHOP}$ = 317 Hz, $P$ = 400 nW, $V_{SD}$ = 100 mV, cw-laser excitation) and background noise without laser excitation (squares). (c) and (d): Chopper frequency dependence of the normalized photoconductance of sample C for parallel and perpendicular polarization at (c) $E_{PHOTON}$ = 1.31 eV and (d) $E_{PHOTON}$ = 1.76 eV, and fit

according to Eq. (1) (line) (at $T_{BATH}$ = 7.2 K, $V_{SD}$ = 100 mV, $P$ = 400 nW, cw-laser excitation). (e) Chopper frequency dependence of the photoconductance of sample B (at $T_{BATH}$ = 3.5 K, $E_{PHOTON}$ = 1.75 eV, $V_{SD}$ = 20 mV, and cw-laser excitation), fit (line) according to Eq. (1), and background noise without laser excitation (squares). (f) Photoconductance of sample D at room temperature ($f_{CHOP}$ = 480 Hz, $P$ = 50 µW, pulsed laser excitation).

FIG. 8: Model for different energy and heat transfer mechanisms within the nanotube-nanocrystal-hybrid and the substrate.

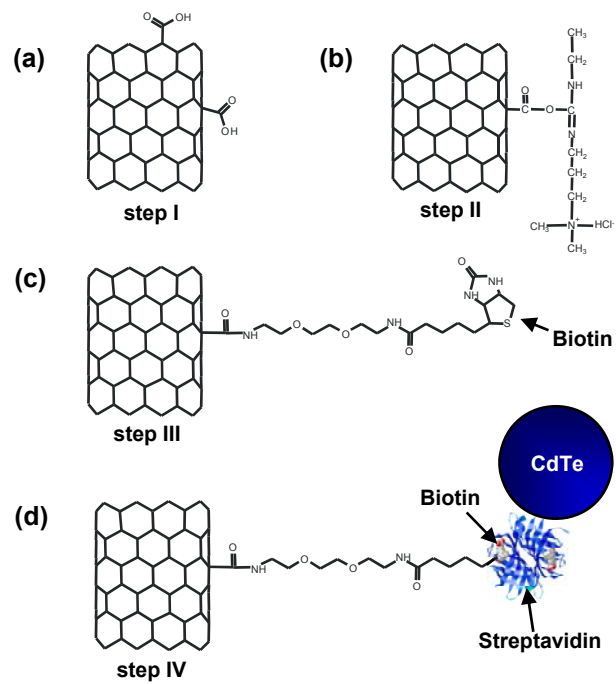

Fig. 1

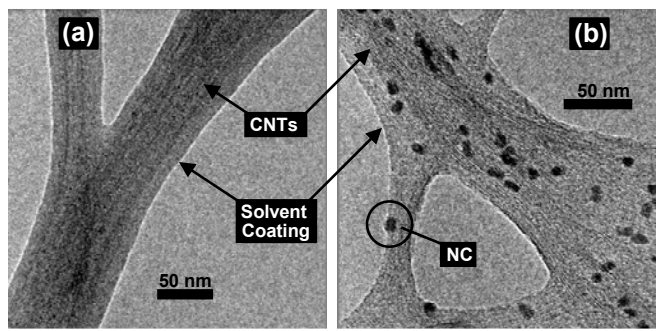

Fig. 2

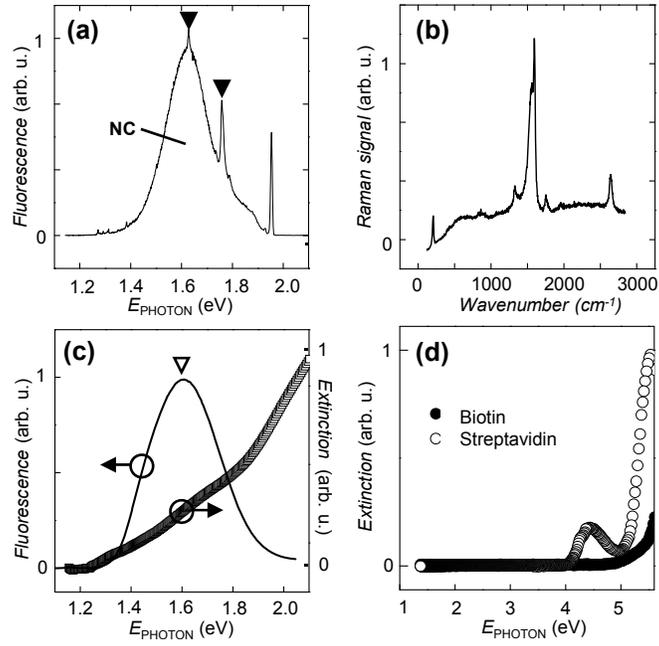

Fig. 3

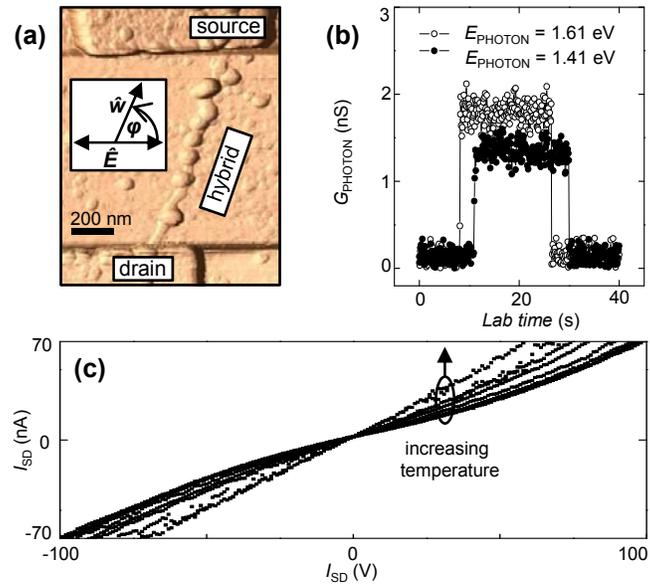

Fig. 4

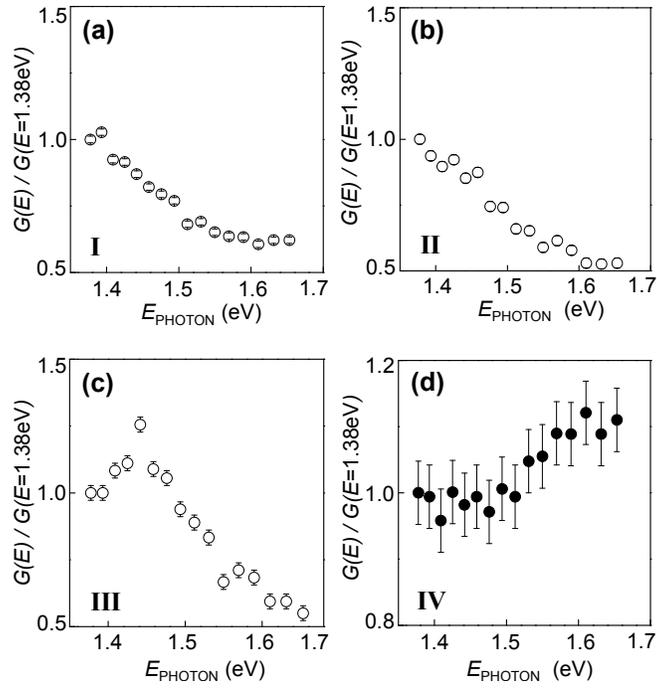

Fig. 5

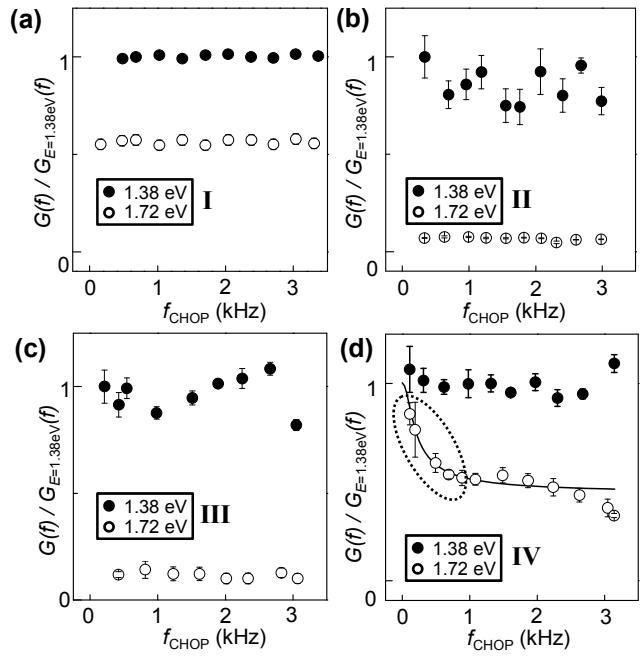

Fig. 6

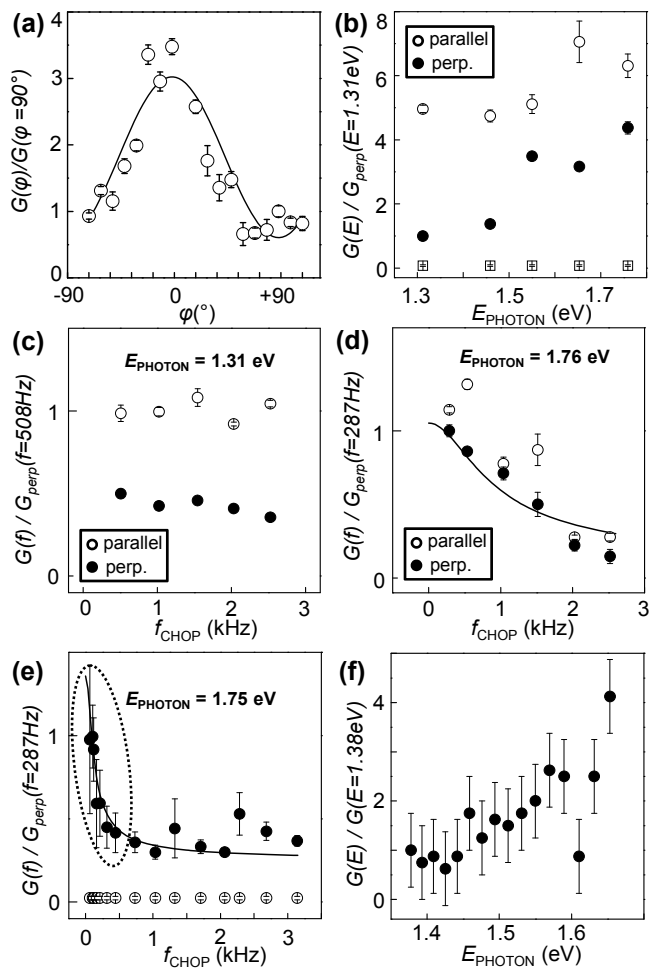

Fig. 7

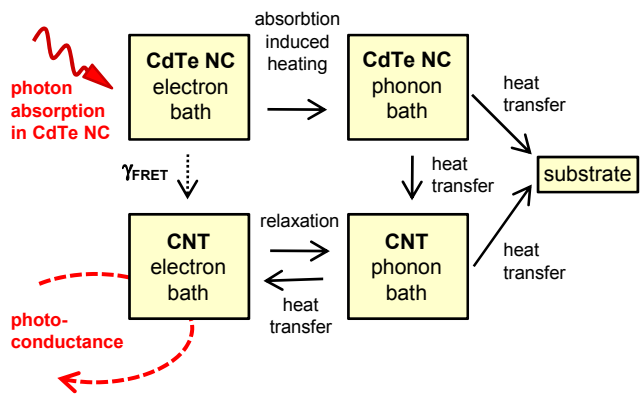

Fig. 8

23